\documentclass[dvipsnames,sigconf]{acmart} 

\usepackage{xcolor}
\usepackage{multirow}
\usepackage{array}
\usepackage{bbm}
\usepackage{float}
\usepackage{subcaption}

\copyrightyear{2020}
\acmYear{2020}
\setcopyright{rightsretained}
\acmConference[ICTIR '20]{Proceedings of the 2020 ACM SIGIR International Conference on the Theory of Information Retrieval}{September 14--17, 2020}{Virtual Event, Norway}
\acmBooktitle{Proceedings of the 2020 ACM SIGIR International Conference on the Theory of Information Retrieval (ICTIR '20), September 14--17, 2020, Virtual Event, Norway}\acmDOI{10.1145/3409256.3409838}
\acmISBN{978-1-4503-8067-6/20/09}

\begin{document}

\title{Understanding BERT Rankers Under Distillation }

\author{Luyu Gao}
\affiliation{%
  \institution{Carnegie Mellon University}
}
\email{luyug@cs.cmu.edu}

\author{Zhuyun Dai}
\affiliation{%
  \institution{Carnegie Mellon University}
}
\email{zhuyund@cs.cmu.edu}

\author{Jamie Callan}
\affiliation{%
  \institution{Carnegie Mellon University}
}
\email{callan@cs.cmu.edu}



\begin{abstract}
Deep language models such as BERT pre-trained on large corpus have given a huge performance boost to the state-of-the-art information retrieval ranking systems. Knowledge embedded in such models allows them to pick up complex matching signals between passages and queries. However, the high computation cost during inference limits their deployment in real-world search scenarios. In this paper, we study if and how the knowledge for search within BERT can be transferred to a smaller ranker through distillation. Our experiments demonstrate that it is crucial to use a proper distillation procedure, which produces up to nine times speedup while preserving the state-of-the-art performance.
\end{abstract}

%


\maketitle

\section{Introduction}
\label{sec-introduction}

Deep language models, such as BERT~\cite{Devlin2019BERTPO} learned from large-scale corpora, have pushed the state-of-the-art of search ranking to a new level. All top-performing teams in the TREC 2019 Deep Learning track used fine-tuned BERT for the final re-ranking stage~\cite{craswell2019overview}. 
Ranking with BERT is effective, but requires computation through multiple transformer layers, which is computationally complex.  We seek a faster model that preserves BERT-based ranking accuracy.

Recent studies suggest that transformer models are over-parame-terized and can be effectively compressed into smaller, faster transformer models through the process of distillation~\cite{Jiao2019TinyBERTDB, Sun2019PatientKD, Sanh2019DistilBERTAD}. 
It is an open question how such distillation affects ranking accuracy.

This paper aims to understand the distillation procedure and its impact on the distilled ranker. We generate distilled rankers of various degrees of compression and with various types of knowledge being distilled. We compare them against the original BERT ranker on a widely-used benchmark dataset to measure changes to ranker accuracy and efficiency. 
We also study the implications of distillations at training time, exploring convergence behaviour during distillation and training time required for convergence.  

The paper's contributions include the following:
\begin{itemize}
    \item We provide a comprehensive evaluation of distilled BERT rankers. To the best of our knowledge, this is the first work to use distillation techniques to improve the efficiency of BERT rankers in information retrieval. 
    \item We investigate various types of knowledge that can be distilled to the ranker. 
    We show that, with a proper distillation approach, a much smaller ranker can be as effective as the state-of-the-art BERT rankers while being an order of magnitude faster. 
    \item We show that  different distillation procedures incur different training cost, and provide recommendations for system developers to trade-off between effectiveness, online evaluation time, and offline training time.
\end{itemize}


\section{Background}
\label{sec-background}


Deep pre-trained language models (LM) are large neural networks trained on surrounding text signals from large text corpora~\cite{Devlin2019BERTPO}. 
These models can then be fine-tuned over other target tasks. Notably, deep LMs such as BERT \cite{Devlin2019BERTPO} have achieved state-of-the-art performance in several natural language tasks, including text search~\cite{Nogueira2019PassageRW, dai2019deeper}.  
In general, BERT rankers are trained by fine-tuning BERT over search logs, using query and passage as the two input sentences and making relevance prediction conditioned on the output sentence/word representations. 
Using hundreds of millions of parameters, BERT learns rich language patterns that are useful for ranking. However, the high complexity makes it computationally expensive to run BERT rankers at a large scale~\cite{nogueira2019document}.

To compress a large neural network, \citet{Hinton2015DistillingTK} propose \emph{distillation}. They use the large network as a teacher to train the small network (student) by minimizing the distance between the two models' output prediction probability distributions. 
 Recently, a family of distillation algorithms have been proposed for distilling large teacher transformers to small student transformers \cite{Sun2019PatientKD,Jiao2019TinyBERTDB}. Compared to \citet{Hinton2015DistillingTK}, they also minimize the distance between student and teacher self attention distributions in the intermediate layers~\cite{transformer}. To the best of our knowledge, there is no prior work studying distilling BERT for search ranking.


 

\section{Distilling BERT for Ranking}
\label{sec-distill}

\begin{figure*}
\centering
    \begin{subfigure}[t]{0.36\textwidth}
    \centering
    \includegraphics[width=0.9\textwidth]{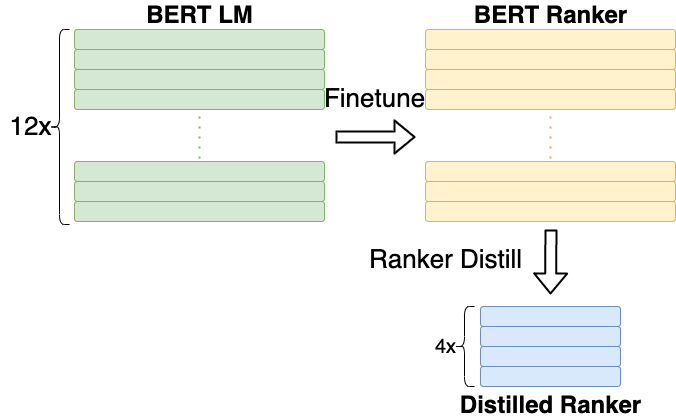}
    \caption{ \centering Ranker Distill}
    \label{fig:distill-rk}
    \end{subfigure}
    \begin{subfigure}[t]{0.29\textwidth}
    \centering
    \includegraphics[width=0.9\textwidth]{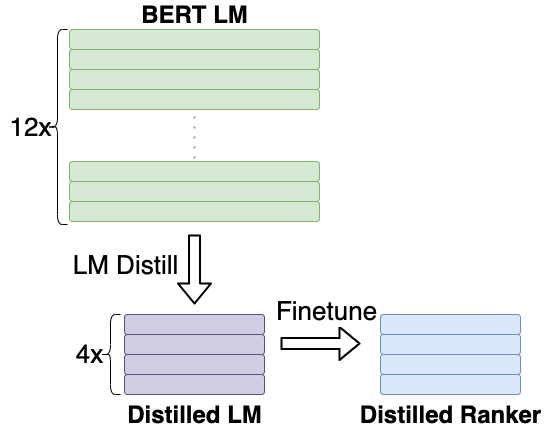}
    \caption{ \centering LM Distill + Fine-tuning}
    \label{fig:distill-lmft}
    \end{subfigure}
       \begin{subfigure}[t]{0.36\textwidth}
    \centering
    \includegraphics[width=0.9\textwidth]{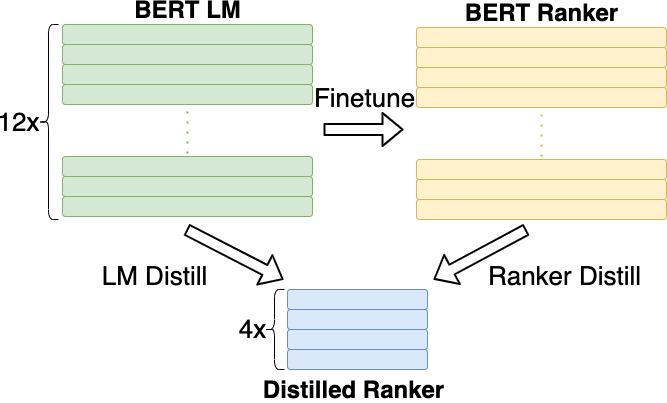}
    \caption{ \centering LM Distill + Ranker Distill}
 \label{fig:distill-lmft}    
 \end{subfigure}
 \caption{Three approaches of using distillation to create a smaller and faster reranker from the larger BERT model. }\label{fig:distill}
\end{figure*}
The power of a BERT ranker is from two main sources: 1)~\emph{general-purpose} language modeling knowledge learned in pre-training, and 2)~\emph{search-specific} relevance modeling knowledge learned in fine-tuning. We desire to produce a smaller and faster ranker also equipped with both types of knowledge. 
Fine-tuning can teach search knowledge, turning an LM into a ranker, while distillation can transfer either LM knowledge (LM Distill) or search knowledge (Ranker Distill) from a large teacher to a small student model.
As shown in Figure~\ref{fig:distill}, here we detail three distinct methods that combine fine-tuning and distillation to arrive at a smaller ranker from an originally full-sized BERT model: 1.~Ranker Distill: distillation is used for search knowledge, 2.~LM Distill + Fine-tuning: distillation is used for LM knowledge, and 3.~LM Distill + Ranker Distill: distillation is used for both LM and search knowledge.



\textbf{Ranker Distill}
The first method distills a BERT ranker's search knowledge to a smaller student ranker, assuming LM knowledge can be learned implicitly.
Before running distillation, this method first fine-tunes BERT LM to generate a BERT ranker. It starts distillation by randomly initializing a student transformer with desired smaller architecture. Then it let BERT ranker to rank documents, and distills the BERT ranker's ranking behavior to the student.

\textbf{LM Distill + Fine-tuning}
The second method teaches general-purpose LM knowledge with distillation, and search-specific knowledge with fine-tuning. It distills a pre-trained BERT language model (BERT LM) to student, assuming it transfers general-purpose LM knowledge enough that the student model can learn search-specific patterns independently through fine-tuning. It first runs a pre-trained BERT over a large text corpora and distill LM predictions from BERT to a smaller transformer, producing a distilled LM. Then, for search-specific knowledge, it treats the distilled LM similarly as the original BERT, and fine-tunes the distilled LM over search logs.

\textbf{LM Distill + Ranker Distill} The third method teaches both general LM knowledge and search-specific knowledge with distillation. 
In particular, it first uses pre-trained BERT LM as the teacher to produce a distilled LM of desired size. Next, a second round of distillation, Ranker Distill, 
proceeds to distill a BERT ranker onto the distilled LM, generating the final ranker.  The LM Distill provides a good initialization, which may improve generalization ability and robustness. The Ranker Distill trains the student ranker with intermediate attention signals as well as output signals from the teacher ranker, providing richer training signals than the straight-forward fine-tuning, which may help the student to perform more similar to the teacher ranker. 

\section{Experiment Methodology}
\label{sec: experiment-method}

\begin{table*}[t]
\caption{Performance of distilled rankers. Time cost refers to the time to generate rankings for one query by reranking the top 1000 documents. 
* indicates non-inferiority to BERT ranker with a 95\% confidence interval \protect \footnotemark .}
\centering
\renewcommand{\arraystretch}{1}
\begin{tabular}{ l || c | c c c | c }
\hline \hline 
\multirow{2}{*}{Method} & \multicolumn{1}{c|}{MARCO Dev Queries} & \multicolumn{3}{c|}{ TREC2019 DL Queries} & \multirow{2}{*}{Time (Speedup)}  \\ 
 & MRR@10 & MRR & NDCG@10 & MAP@1000 &  \\  
 \hline
 BERT ranker (12 layers) & 0.3527 & 0.9349 & 0.7032 & 0.4836 & \multirow{1}{*}{2.97s (1.0$\times$)} \\
 \hline
 6-layer distilled ranker & & & & & \multirow{4}{*}{1.50s (2.0$\times$)}\\

 Ranker Distill  & 0.3380 & 0.9271 & 0.6856 & 0.4828 & \\
  LM Distill + Fine-Tuning & 0.3556* &0.9651* & 0.7191* & 0.4918* &  \\
 LM Distill + Ranker Distill & 0.3600*  & 0.9516* & 0.6923 & 0.4924* &   \\
 \hline
 4-layer distilled rankers & &  & &  & \multirow{4}{*}{0.33s (9.0$\times$)}\\
 Ranker Distill & 0.3286 & 0.9351 & 0.6693 & 0.4589 &  \\
  LM Distill + Fine-Tuning & 0.3320 &  0.9496* & 0.6812 & 0.4609 & \\
  LM Distill + Ranker Distill & 0.3501* & 0.9291*	& 0.6827* & 0.4819* & \\
 \hline \hline 
\end{tabular}
\label{tab:perf}
\end{table*}


\textbf{Dataset} We use the MS MARCO Passage Ranking task dataset~\cite{nguyen2016ms}. MS MARCO contains around 8.8 million passages and around $0.5$ million real queries with judgments as training data. All BERT rankers are tested on a reranking task where the model is asked  to rerank the top 1000 documents retrieved by BM25. Two evaluation query sets with different characteristics are used in this work. 

\textbf{MS MARCO Dev Queries}: this evaluation query set contains 6980 queries from MS MARCO dataset's development set, which has been widely used in prior research~\cite{Nogueira2019PassageRW}. Most of the queries have only one document judged relevant; the relevance labels are binary. Following~\cite{nguyen2016ms}, we used MRR@10 to evaluate the ranking accuracy on this query set.

\textbf{TREC2019 DL Queries}: this evaluation query set is the official evaluation query set used in the TREC 2019 Deep Learning Track~\cite{craswell2019overview}. It contains 43 queries with multiple relevant documents manually judged by NIST assessors with graded relevance labels. On average, a query has 95 relevant documents.
Following~\citet{craswell2019overview}, we used MRR, NDCG@10, and MAP@1000 to evaluate the ranking accuracy on this query set.

\textbf{Models for Comparison} The BERT ranker serves as our main performance baseline.  We train a BERT ranker by fine-tuning the BERT \textit{base-uncased} model over MS MARCO's training set following~\citet{Nogueira2019PassageRW}.

For distilled models, we investigate two different student architectures to study impacts from applying different degrees of compression.
The first, the \textbf{6-layer} model, is a six layer model with transformer hidden dimension of 768. The second, the \textbf{4-layer} model, is a four layer model with hidden dimension of 312. The settings are similar to those studied in \cite{Jiao2019TinyBERTDB}. 
Combining the three distillation methods discussed in Section~\ref{sec-distill} ( Ranker Distill, LM Distill + Fine-tuning, and LM Distill + Ranker Distill) and the two aforementioned architectures (6-layer, 4-layer), we use a total of six types of distilled rankers in the experiment. 

The implementation of model distillation was based on the TinyBERT software~\cite{Jiao2019TinyBERTDB}. LM Distill uses the uncased Bert-base model~\cite{Devlin2019BERTPO} as the teacher model. For Ranker Distill, the teacher model is the baseline BERT ranker fine-tuned on MS MARCO.
We use Adam optimizer with a weight decay of 0.01 and a learning rate of 2e-5 for distillation. We use AdamW optimizer with a learning rate 2e-5 for fine-tuning both BERT ranker and distilled ranker. Training is done on 4 RTX 2080 TI GPU, while inference on 1 GPU. 

\section{Experiment Results}
\label{sec: experiment-result}

Two experiments study the distilled ranker's performance during inference, and distillation's implications during training. 
 
\subsection{Impacts on Ranking Performance}

The first set of experiments aims to understand, \emph{can distillation speed up a BERT reranker while retaining its effectiveness?} To answer this question, we study the various distilled models' performance during inference time, including their effectiveness, efficiency, and robustness to various reranking depth.

\textbf{ Effectiveness and Efficiency} Table~\ref{tab:perf} reports the ranking accuracy of various distilled rankers. We evaluate the rankers at training checkpoints 10K, 100K, 1M, 5M, 10M. For each model, we report the best performance among all checkpoints. 

We found Ranker Distill gives the worst performance. LM Distill + Fine-tuning is able to reach original BERT ranker's effectiveness with a 6-layer distilled ranker, but fails with a 4-layer distilled ranker. LM Distill + Ranker Distill yields the strongest performance, being non-inferior to original BERT ranker in both cases with statistical significance.  
Observation is consistent across two evaluation sets. 

Table~\ref{tab:perf} also reports the rankers' speed measured by the time needed to rank one query with 1,000 candidate documents. We test with feeding in data with batch size 64, 128, 256 and 512 to the GPU and record the number of the most efficient batch size. As can be seen, the 6-layer configuration can be 2 times faster than the full BERT ranker, and the 4-layer distilled rankers can achieve 9 times speedup, thanks to reduction in both model dimension as well as number of model layer. Importantly, it demonstrates that a properly distillation procedure (LM Disitll + Ranker Distill) can compress the BERT ranker into much smaller ones \emph{without hurting effectiveness while being 9 times faster}. 

\footnotetext{The equivalence is established by rejecting the null hypothesis that distilled ranker is at least 3\% worse than original BERT ranker with a 95\% confidence interval.}

\textbf{Effects of Different Distillation Procedures} 
Rank Distill straight-forwardly distills a fine-tuned BERT onto a randomly initialized student model. However, based on our results, this is not sufficient for the distilled ranker to recover all teacher model's effectiveness. Using Rank Distill alone, the distilled ranker learns everything from the fine-tuned BERT ranker, without explicitly learning general-purpose language knowledge. On the other hand, the two approaches using LM Distill achieve substantially higher performance than Rank Distill, demonstrating that it is critical for the distilled ranker to learn general-purpose language modeling knowledge through LM Distill explicitly. 

The  LM Distll + Fine-tuning method achieves similar results as the original BERT ranker when using relatively large models (6-layer), indicating that it is possible to directly fine-tune a distilled LM for downstream ranking tasks. However, when using a smaller model (4-layer), the accuracy of a fine-tuned distilled LM is slightly lower than a fine-tuned BERT. A higher degree of compression may lose too much LM knowledge, hurting the model's ability to adapt to the downstream ranking task. On the contrary, LM Distill + Ranker Distill can be equally accurate as the original BERT ranker using a small model (4-layer). Ranker Distill  provides richer training signals than the fine-tuning method, helping the student to recover the teacher's effectiveness even under a high compression rate.

To summarize, our results indicate that to distill a BERT ranker effectively, it is critical to explicitly distill the general-purpose language  modeling  knowledge first through LM Distill. The search-specific knowledge can be learned from direct fine-tuning or Ranker Distill, depending on the desired model size -- while a simple fine-tuning is sufficient for larger distilled models, smaller models still need Ranker Distill to be as effective as the original BERT ranker.

\textbf{Robustness to Reranking Depth} We also investigated the distilled reranker's performance at various reranking depth. Table~\ref{tab:depth-perf} shows the results of using our best distilled ranker (LM Distill + Rank Distill) to rerank the top 10 to 1000 passages retrieved by BM25. 
As shown in Table~\ref{tab:depth-perf}, various models' performance is  consistent across all reranking depth, further confirming our finding that distillation effectively compresses the model. Furthermore, dropping the reranking depth to 100 results in only 4\% drop in MRR@10. Based on our speed measurement in table \ref{tab:perf}, this implies that in case where 4\% of performance drop is acceptable, a 4-layer distilled ranker that runs at the speed of $0.03s$ per query, processing 30 queries per second on a single GPU machine.

\begin{table}[t]
\caption{Robustness of distilled ranker to various reranking depths. Distilled ranker used LM Distill + Ranker Distill, the best configuration found in Table~\ref{tab:perf}.}
\renewcommand{\arraystretch}{1}
\begin{tabular}{ m{2cm} | c c c}
\hline
 Depth & BERT ranker & 6-layer & 4-layer \\  
 \hline
 10 &  0.2716 & 0.2750 & 0.2717  \\
 20 & 0.2978 & 0.3035 & 0.2979\\
 50 & 0.3237 & 0.3303 & 0.3247 \\
 100 & 0.3367 & 0.3440 & 0.3376 \\
 200 & 0.3436 & 0.3516 & 0.3437\\
 1000 & 0.3527 & 0.3600 & 0.3501 \\
 \hline
\end{tabular}
\label{tab:depth-perf}
\end{table}


\subsection{Impacts on Training}
The previous section shows how distillation affects rankers' effectiveness and efficiency at \emph{inference} time. This section discusses how distilled rankers behave at \emph{training} time. As our experiment shows that Ranker Distill alone is not sufficient, here we focus on the training of LM Distill + Fine-tuning and LM Distill + Ranker Distill. 

\begin{figure}[t]
    \centering
    \includegraphics[scale=0.38]{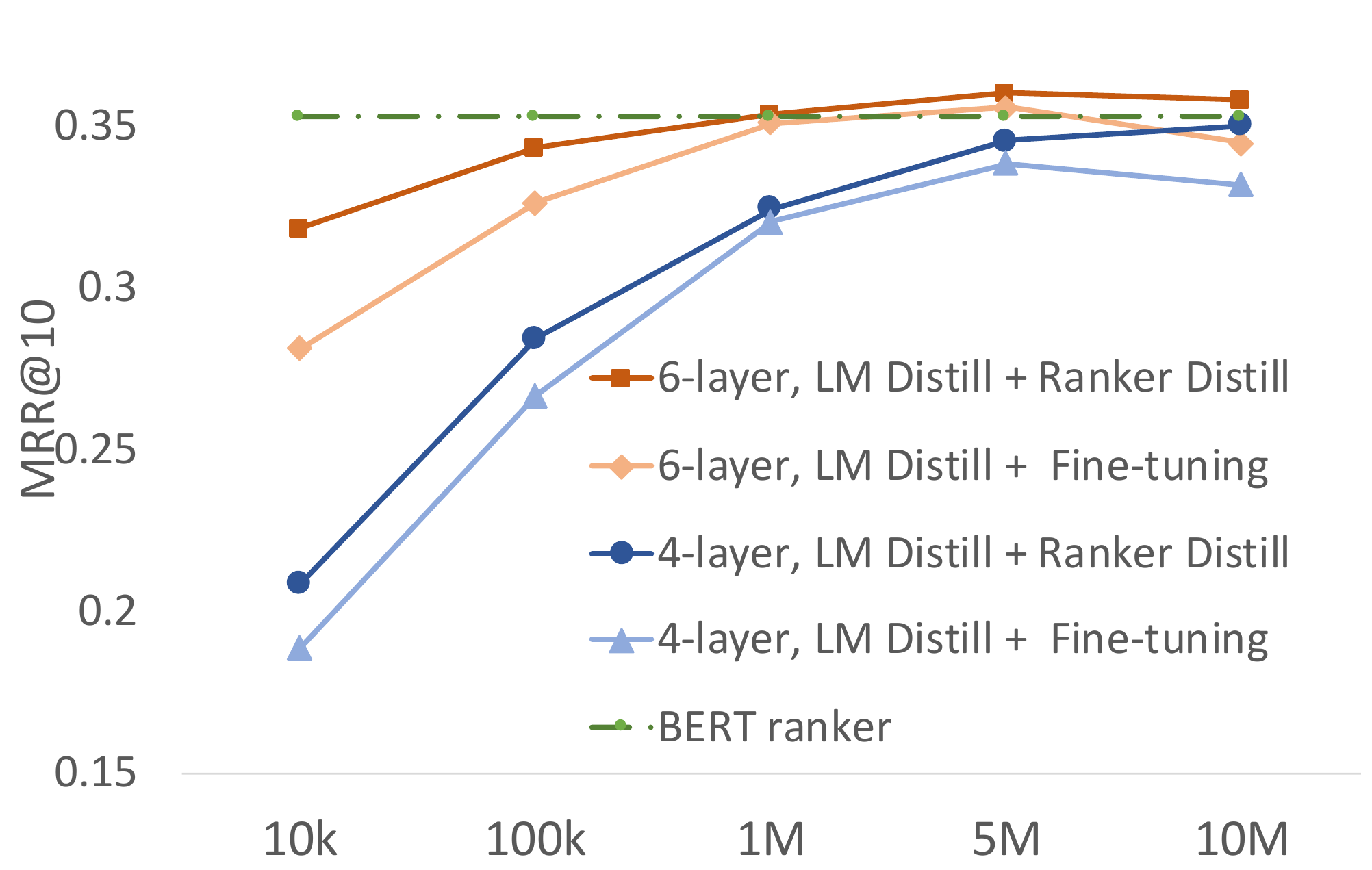}
    \caption{Effect of Training Data Amount}
    \label{fig:data-amount}
\end{figure}

\textbf{Effects of Model Size on Convergence} We initialize the distilled ranker with LM Distill, and train with Ranker Distill or fine-tuning using 10K, 100K, 1M, 5M and 10M of training examples. We plot trend of MRR@10 in Figure \ref{fig:data-amount}.  It shows a clear distinction between the smaller model (4-layer) and the bigger model (6-layer). The 6-layer model converges quickly, achieving a reasonable performance at 10K training pairs, and close to the full BERT ranker with roughly 100K pairs. Meanwhile, the 4-layer model converges slower, requiring 5M to 10M pairs to be close to the full BERT ranker. Based on these results, we conclude that smaller distilled models are overall more data-hungry due to the loss of LM knowledge.

\textbf{Fine-tuning vs. Ranker Distill during Training} As discussed previously, LM Distill + Fine-tuning and LM Distll + Ranker Distill can both generate reasonably effective rankers, while the latter is more effective in compression.
This experiment aims to further understand their training behavior.  
As shown in Figure~\ref{fig:data-amount}, the same model consume less data to converge with distillation than with fine-tuning. 
Distillation also generalizes better:
accuracy of fine-tuning starts to decrease after training on 5M query-document pairs, indicating potential
overfitting. Meanwhile, the accuracy of Ranker Distill has not plateaued yet, showing that for a small model, learning search-specific knowledge through distillation is less prone to overfitting than directly learning from fine-tuning.


Ranker Distill and Fine-tuning also differ in \emph{amount of training time}. 
Ranker Distill runs both teacher model and student model, leading to higher compute and memory cost.
Table~\ref{tab:runtime} shows time it takes for various model to reach close to (within 5\%) BERT ranker's performance. 
As shown in Table~\ref{tab:runtime}, fine-tuning is 5 to 10 times faster than Ranker Distill when training.
The 4-layer model, though faster in inference, trains slower time and takes 5 times more training steps to reach close to BERT ranker's performance (Figure~\ref{fig:data-amount}). 

\par
\begin{table}[t]
\caption{Amount of training time for distilled models to reach the first checkpoint that is close to BERT ranker's performance, i.e., 1M training examples for the 6-layer model and 5M for the 4-layer model. }
\renewcommand{\arraystretch}{0.9}

\begin{tabular}{ l | c c }
\hline
 Model & 6-layer & 4-layer \\  
 \hline
  Fine-Tuning  & 1.1h  & 1.6h  \\
 Ranker Distill   & 5.5h & 14.8h \\
 \hline
\end{tabular}
\label{tab:runtime}
\end{table}
To summarize, highly compressed rankers require more training data, longer training time, and may fail to converge to best optimum with fine-tuning.
One should choose distillation approaches by weighing importance of off-line training time, online performance, and online response time. When ranker requires no update, LM Distill + Rank Distill with small models offers the best online-serving speed. When the ranker needs to be frequently updated, doing LM Distill + Fine-tuning is generally preferred. 





\section{Conclusion}
\label{sec-conclusion}
In this paper, we demonstrate that, with distillation, one can turn a BERT ranker into a substantially faster ranker, while still preserving the BERT ranker's ranking performance. We evaluated different ways to generate distilled rankers from the original BERT LM and identified that the most robust and effective method is a two-round distillation where both the general-purpose LM knowledge and search-specific knowledge are distilled. 
In comparison, removal of each leads to an inferior ranker.
We also examine distilled models' training behaviors. We found a higher degree of compression introduces more difficulties in model optimization due to loss of knowledge: though faster in inference, smaller distilled models consume more training data and longer training time.    

\section*{Acknowledgments}

This work was supported by the National Science Foundation (NSF) grant IIS-1815528 and The Boeing Company.
Any opinions, findings, and conclusions in this paper are the authors' and do not necessarily reflect those of the sponsors.

\bibliographystyle{ACM-Reference-Format}
\bibliography{strings-short, local}


\begin{thebibliography}{11}


\ifx \showCODEN    \undefined \def \showCODEN     #1{\unskip}     \fi
\ifx \showDOI      \undefined \def \showDOI       #1{#1}\fi
\ifx \showISBNx    \undefined \def \showISBNx     #1{\unskip}     \fi
\ifx \showISBNxiii \undefined \def \showISBNxiii  #1{\unskip}     \fi
\ifx \showISSN     \undefined \def \showISSN      #1{\unskip}     \fi
\ifx \showLCCN     \undefined \def \showLCCN      #1{\unskip}     \fi
\ifx \shownote     \undefined \def \shownote      #1{#1}          \fi
\ifx \showarticletitle \undefined \def \showarticletitle #1{#1}   \fi
\ifx \showURL      \undefined \def \showURL       {\relax}        \fi
\providecommand\bibfield[2]{#2}
\providecommand\bibinfo[2]{#2}
\providecommand\natexlab[1]{#1}
\providecommand\showeprint[2][]{arXiv:#2}

\bibitem[\protect\citeauthoryear{Craswell, Mitra, Yilmaz, and Campos}{Craswell
  et~al\mbox{.}}{2019}]%
        {craswell2019overview}
\bibfield{author}{\bibinfo{person}{Nick Craswell}, \bibinfo{person}{Bhaskar
  Mitra}, \bibinfo{person}{Emine Yilmaz}, {and} \bibinfo{person}{Daniel
  Campos}.} \bibinfo{year}{2019}\natexlab{}.
\newblock \showarticletitle{Overview of the TREC 2019 deep learning track}. In
  \bibinfo{booktitle}{\emph{TREC (to appear)}}.
\newblock


\bibitem[\protect\citeauthoryear{Dai and Callan}{Dai and Callan}{2019}]%
        {dai2019deeper}
\bibfield{author}{\bibinfo{person}{Zhuyun Dai} {and} \bibinfo{person}{Jamie
  Callan}.} \bibinfo{year}{2019}\natexlab{}.
\newblock \showarticletitle{Deeper Text Understanding for IR with Contextual
  Neural Language Modeling}. In \bibinfo{booktitle}{\emph{The 42nd
  International ACM SIGIR Conference on Research \& Development in Information
  Retrieval}}.
\newblock


\bibitem[\protect\citeauthoryear{Devlin, Chang, Lee, and Toutanova}{Devlin
  et~al\mbox{.}}{2019}]%
        {Devlin2019BERTPO}
\bibfield{author}{\bibinfo{person}{Jacob Devlin}, \bibinfo{person}{Ming-Wei
  Chang}, \bibinfo{person}{Kenton Lee}, {and} \bibinfo{person}{Kristina
  Toutanova}.} \bibinfo{year}{2019}\natexlab{}.
\newblock \bibinfo{title}{BERT: Pre-training of Deep Bidirectional Transformers
  for Language Understanding}.
\newblock
\newblock


\bibitem[\protect\citeauthoryear{Hinton, Vinyals, and Dean}{Hinton
  et~al\mbox{.}}{2015}]%
        {Hinton2015DistillingTK}
\bibfield{author}{\bibinfo{person}{Geoffrey~E. Hinton}, \bibinfo{person}{Oriol
  Vinyals}, {and} \bibinfo{person}{Jeffrey Dean}.}
  \bibinfo{year}{2015}\natexlab{}.
\newblock \showarticletitle{Distilling the Knowledge in a Neural Network}.
\newblock \bibinfo{journal}{\emph{ArXiv}}  \bibinfo{volume}{abs/1503.02531}
  (\bibinfo{year}{2015}).
\newblock


\bibitem[\protect\citeauthoryear{Jiao, Yin, Shang, Jiang, Chen, Li, Wang, and
  Liu}{Jiao et~al\mbox{.}}{2019}]%
        {Jiao2019TinyBERTDB}
\bibfield{author}{\bibinfo{person}{Xiaoqi Jiao}, \bibinfo{person}{Y. Yin},
  \bibinfo{person}{Lifeng Shang}, \bibinfo{person}{Xin Jiang},
  \bibinfo{person}{Xusong Chen}, \bibinfo{person}{Linlin Li},
  \bibinfo{person}{Fang Wang}, {and} \bibinfo{person}{Qun Liu}.}
  \bibinfo{year}{2019}\natexlab{}.
\newblock \showarticletitle{TinyBERT: Distilling BERT for Natural Language
  Understanding}.
\newblock \bibinfo{journal}{\emph{ArXiv}}  \bibinfo{volume}{abs/1909.10351}
  (\bibinfo{year}{2019}).
\newblock


\bibitem[\protect\citeauthoryear{Nguyen, Rosenberg, Song, Gao, Tiwary,
  Majumder, and Deng}{Nguyen et~al\mbox{.}}{2016}]%
        {nguyen2016ms}
\bibfield{author}{\bibinfo{person}{Tri Nguyen}, \bibinfo{person}{Mir
  Rosenberg}, \bibinfo{person}{Xia Song}, \bibinfo{person}{Jianfeng Gao},
  \bibinfo{person}{Saurabh Tiwary}, \bibinfo{person}{Rangan Majumder}, {and}
  \bibinfo{person}{Li Deng}.} \bibinfo{year}{2016}\natexlab{}.
\newblock \showarticletitle{MS MARCO: A human generated machine reading
  comprehension dataset}.
\newblock \bibinfo{journal}{\emph{arXiv preprint arXiv:1611.09268}}
  (\bibinfo{year}{2016}).
\newblock


\bibitem[\protect\citeauthoryear{Nogueira and Cho}{Nogueira and Cho}{2019}]%
        {Nogueira2019PassageRW}
\bibfield{author}{\bibinfo{person}{Rodrigo Nogueira} {and}
  \bibinfo{person}{Kyunghyun Cho}.} \bibinfo{year}{2019}\natexlab{}.
\newblock \showarticletitle{Passage Re-ranking with BERT}.
\newblock \bibinfo{journal}{\emph{ArXiv}}  \bibinfo{volume}{abs/1901.04085}
  (\bibinfo{year}{2019}).
\newblock


\bibitem[\protect\citeauthoryear{Nogueira, Yang, Lin, and Cho}{Nogueira
  et~al\mbox{.}}{2019}]%
        {nogueira2019document}
\bibfield{author}{\bibinfo{person}{Rodrigo Nogueira}, \bibinfo{person}{Wei
  Yang}, \bibinfo{person}{Jimmy Lin}, {and} \bibinfo{person}{Kyunghyun Cho}.}
  \bibinfo{year}{2019}\natexlab{}.
\newblock \showarticletitle{Document expansion by query prediction}.
\newblock \bibinfo{journal}{\emph{arXiv preprint arXiv:1904.08375}}
  (\bibinfo{year}{2019}).
\newblock


\bibitem[\protect\citeauthoryear{Sanh, Debut, Chaumond, and Wolf}{Sanh
  et~al\mbox{.}}{2019}]%
        {Sanh2019DistilBERTAD}
\bibfield{author}{\bibinfo{person}{Victor Sanh}, \bibinfo{person}{Lysandre
  Debut}, \bibinfo{person}{Julien Chaumond}, {and} \bibinfo{person}{Thomas
  Wolf}.} \bibinfo{year}{2019}\natexlab{}.
\newblock \showarticletitle{DistilBERT, a distilled version of BERT: smaller,
  faster, cheaper and lighter}.
\newblock \bibinfo{journal}{\emph{ArXiv}}  \bibinfo{volume}{abs/1910.01108}
  (\bibinfo{year}{2019}).
\newblock


\bibitem[\protect\citeauthoryear{Sun, Cheng, Gan, and Liu}{Sun
  et~al\mbox{.}}{2019}]%
        {Sun2019PatientKD}
\bibfield{author}{\bibinfo{person}{Siqi Sun}, \bibinfo{person}{Yu Cheng},
  \bibinfo{person}{Zhe Gan}, {and} \bibinfo{person}{Jingjing Liu}.}
  \bibinfo{year}{2019}\natexlab{}.
\newblock \showarticletitle{Patient Knowledge Distillation for BERT Model
  Compression}. In \bibinfo{booktitle}{\emph{EMNLP/IJCNLP}}.
\newblock


\bibitem[\protect\citeauthoryear{Vaswani, Shazeer, Parmar, Uszkoreit, Jones,
  Gomez, Kaiser, and Polosukhin}{Vaswani et~al\mbox{.}}{2017}]%
        {transformer}
\bibfield{author}{\bibinfo{person}{Ashish Vaswani}, \bibinfo{person}{Noam
  Shazeer}, \bibinfo{person}{Niki Parmar}, \bibinfo{person}{Jakob Uszkoreit},
  \bibinfo{person}{Llion Jones}, \bibinfo{person}{Aidan~N. Gomez},
  \bibinfo{person}{Lukasz Kaiser}, {and} \bibinfo{person}{Illia Polosukhin}.}
  \bibinfo{year}{2017}\natexlab{}.
\newblock \showarticletitle{Attention is All you Need}.
\newblock \bibinfo{journal}{\emph{ArXiv}}  \bibinfo{volume}{abs/1706.03762}
  (\bibinfo{year}{2017}).
\newblock


\end{thebibliography}








\end{document}